\input harvmac
\input amssym

\def\unit{\relax{\rm 1\kern-.26em I}}
\def\nada{\relax{\rm 0\kern-.30em l}}
\def\tilde{\widetilde}



\noblackbox
\def\IL{\relax{\rm I\kern-.18em L}}
\def\IH{\relax{\rm I\kern-.18em H}}
\def\IR{\relax{\rm I\kern-.18em R}}
\def\IC{\relax\hbox{$\inbar\kern-.3em{\rm C}$}}
\def\IZ{\relax\ifmmode\mathchoice
{\hbox{\cmss Z\kern-.4em Z}}{\hbox{\cmss Z\kern-.4em Z}}
{\lower.9pt\hbox{\cmsss Z\kern-.4em Z}} {\lower1.2pt\hbox{\cmsss
Z\kern-.4em Z}}\else{\cmss Z\kern-.4em Z}\fi}

\def\CJ {{\cal J}}

\def\CL {{\cal L}}


\font\manual=manfnt \def\dbend{\lower3.5pt\hbox{\manual\char127}}

\def\IZ{\relax\ifmmode\mathchoice
{\hbox{\cmss Z\kern-.4em Z}}{\hbox{\cmss Z\kern-.4em Z}}
{\lower.9pt\hbox{\cmsss Z\kern-.4em Z}} {\lower1.2pt\hbox{\cmsss
Z\kern-.4em Z}}\else{\cmss Z\kern-.4em Z}\fi}
\def\half {{1\over 2}}

\def\rt2{\sqrt{2}}
\def\irt2{{1\over\sqrt{2}}}

\def\slashchar#1{\setbox0=\hbox{$#1$}           
   \dimen0=\wd0                                 
   \setbox1=\hbox{/} \dimen1=\wd1               
   \ifdim\dimen0>\dimen1                        
      \rlap{\hbox to \dimen0{\hfil/\hfil}}      
      #1                                        
   \else                                        
      \rlap{\hbox to \dimen1{\hfil$#1$\hfil}}   
      /                                         
   \fi}

\def\foursqr#1#2{{\vcenter{\vbox{
    \hrule height.#2pt
    \hbox{\vrule width.#2pt height#1pt \kern#1pt
    \vrule width.#2pt}
    \hrule height.#2pt
    \hrule height.#2pt
    \hbox{\vrule width.#2pt height#1pt \kern#1pt
    \vrule width.#2pt}
    \hrule height.#2pt
        \hrule height.#2pt
    \hbox{\vrule width.#2pt height#1pt \kern#1pt
    \vrule width.#2pt}
    \hrule height.#2pt
        \hrule height.#2pt
    \hbox{\vrule width.#2pt height#1pt \kern#1pt
    \vrule width.#2pt}
    \hrule height.#2pt}}}}
\def\psqr#1#2{{\vcenter{\vbox{\hrule height.#2pt
    \hbox{\vrule width.#2pt height#1pt \kern#1pt
    \vrule width.#2pt}
    \hrule height.#2pt \hrule height.#2pt
    \hbox{\vrule width.#2pt height#1pt \kern#1pt
    \vrule width.#2pt}
    \hrule height.#2pt}}}}
\def\sqr#1#2{{\vcenter{\vbox{\hrule height.#2pt
    \hbox{\vrule width.#2pt height#1pt \kern#1pt
    \vrule width.#2pt}
    \hrule height.#2pt}}}}

\def\figin{\epsfcheck\figin}\def\figins{\epsfcheck\figins}
\def\epsfcheck{\ifx\epsfbox\UnDeFiNeD
\message{(NO epsf.tex, FIGURES WILL BE IGNORED)}
\gdef\figin##1{\vskip2in}\gdef\figins##1{\hskip.5in}
\else\message{(FIGURES WILL BE INCLUDED)}%
\gdef\figin##1{##1}\gdef\figins##1{##1}\fi}
\def\DefWarn#1{}
\def\figinsert{\goodbreak\midinsert}
\def\ifig#1#2#3{\DefWarn#1\xdef#1{fig.~\the\figno}
\writedef{#1\leftbracket fig.\noexpand~\the\figno}%
\figinsert\figin{\centerline{#3}}\medskip\centerline{\vbox{\baselineskip12pt
\advance\hsize by -1truein\noindent\footnotefont{\bf
Fig.~\the\figno:\ } \it#2}}
\bigskip\endinsert\global\advance\figno by1}

\lref\GeorgiGP{
  H.~Georgi,
  ``New Realization of Chiral Symmetry,''
  Phys.\ Rev.\ Lett.\  {\bf 63}, 1917 (1989).
}

\lref\GeorgiXY{
  H.~Georgi,
  ``Vector Realization Of Chiral Symmetry,''
  Nucl.\ Phys.\  B {\bf 331}, 311 (1990).
}

\lref\KarchPV{
  A.~Karch, E.~Katz, D.~T.~Son and M.~A.~Stephanov,
  ``Linear Confinement and AdS/QCD,''
  Phys.\ Rev.\  D {\bf 74}, 015005 (2006)
  [arXiv:hep-ph/0602229].
}

\lref\ErlichQH{
  J.~Erlich, E.~Katz, D.~T.~Son and M.~A.~Stephanov,
  ``QCD and a Holographic Model of Hadrons,''
  Phys.\ Rev.\ Lett.\  {\bf 95}, 261602 (2005)
  [arXiv:hep-ph/0501128].
}

\lref\SonET{
  D.~T.~Son and M.~A.~Stephanov,
  ``QCD and dimensional deconstruction,''
  Phys.\ Rev.\  D {\bf 69}, 065020 (2004)
  [arXiv:hep-ph/0304182].
}

\lref\HongNP{
  S.~Hong, S.~Yoon and M.~J.~Strassler,
  ``On the couplings of the rho meson in AdS/QCD,''
  arXiv:hep-ph/0501197.
}

\lref\HongSA{
  S.~Hong, S.~Yoon and M.~J.~Strassler,
 ``On the couplings of vector mesons in AdS/QCD,''
  JHEP {\bf 0604}, 003 (2006)
  [arXiv:hep-th/0409118].
}

\lref\WittenKH{
  E.~Witten,
  ``Baryons In The 1/N Expansion,''
  Nucl.\ Phys.\  B {\bf 160}, 57 (1979).
}

\lref\WeinbergKQ{
  S.~Weinberg and E.~Witten,
  ``Limits On Massless Particles,''
  Phys.\ Lett.\  B {\bf 96}, 59 (1980).
}

\lref\BirseHD{
  M.~C.~Birse,
  ``Effective chiral Lagrangians for spin-1 mesons,''
  Z.\ Phys.\  A {\bf 355}, 231 (1996)
  [arXiv:hep-ph/9603251].
}

\lref\WeinbergKR{
  S.~Weinberg,
  ``The quantum theory of fields. Vol. 2: Modern applications,''
{\it  Cambridge, UK: Univ. Pr. (1996) 489 p}
}

\lref\WittenVV{
  E.~Witten,
  ``Current Algebra Theorems For The U(1) Goldstone Boson,''
  Nucl.\ Phys.\  B {\bf 156}, 269 (1979).
}

\lref\WeinbergKJ{
  S.~Weinberg,
  ``Precise relations between the spectra of vector and axial vector mesons,''
  Phys.\ Rev.\ Lett.\  {\bf 18}, 507 (1967).
}

\lref\BernardCD{
  C.~W.~Bernard, A.~Duncan, J.~LoSecco and S.~Weinberg,
  ``Exact Spectral Function Sum Rules,''
  Phys.\ Rev.\  D {\bf 12}, 792 (1975).
}

\lref\DonoghueXB{
  J.~F.~Donoghue and E.~Golowich,
  ``Chiral sum rules and their phenomenology,''
  Phys.\ Rev.\  D {\bf 49}, 1513 (1994)
  [arXiv:hep-ph/9307262].
}

\lref\KawarabayashiKD{
  K.~Kawarabayashi and M.~Suzuki,
  ``Partially conserved axial vector current and the decays of vector mesons,''
  Phys.\ Rev.\ Lett.\  {\bf 16}, 255 (1966).
}

\lref\RiazuddinSW{
  Riazuddin and Fayyazuddin,
  ``Algebra of current components and decay widths of rho and K* mesons,''
  Phys.\ Rev.\  {\bf 147}, 1071 (1966).
}

\lref\ShifmanBX{
  M.~A.~Shifman, A.~I.~Vainshtein and V.~I.~Zakharov,
  ``QCD And Resonance Physics. Sum Rules,''
  Nucl.\ Phys.\  B {\bf 147}, 385 (1979).
}

\lref\HornTM{
  T.~Horn {\it et al.}  [Jefferson Lab F(pi)-2 Collaboration],
  ``Determination of the Charged Pion Form Factor at Q2=1.60 and 2.45
  (GeV/c)2,''
  Phys.\ Rev.\ Lett.\  {\bf 97}, 192001 (2006)
  [arXiv:nucl-ex/0607005].
}

\lref\HuberID{
  G.~M.~Huber {\it et al.}  [Jefferson Lab Collaboration],
  ``Charged pion form factor between $Q^2=0.60$ and $2.45$ GeV$^2$. II. Determination
  of, and results for, the pion form factor,''
  Phys.\ Rev.\  C {\bf 78}, 045203 (2008)
  [arXiv:0809.3052 [nucl-ex]].
}

\lref\tHooftJZ{
  G.~'t Hooft,
  ``A Planar Diagram Theory for Strong Interactions,''
  Nucl.\ Phys.\  B {\bf 72}, 461 (1974).
}

\lref\KweeDD{
  H.~J.~Kwee and R.~F.~Lebed,
  ``Pion Form Factors in Holographic QCD,''
  JHEP {\bf 0801}, 027 (2008)
  [arXiv:0708.4054 [hep-ph]].
}

\lref\RodriguezGomezZP{
  D.~Rodriguez-Gomez and J.~Ward,
  ``Electromagnetic form factors from the fifth dimension,''
  JHEP {\bf 0809}, 103 (2008)
  [arXiv:0803.3475 [hep-th]].
}

\lref\KweeNQ{
  H.~J.~Kwee and R.~F.~Lebed,
  ``Pion Form Factor in Improved Holographic QCD Backgrounds,''
  Phys.\ Rev.\  D {\bf 77}, 115007 (2008)
  [arXiv:0712.1811 [hep-ph]].
}

\lref\HongDQ{
  D.~K.~Hong, M.~Rho, H.~U.~Yee and P.~Yi,
  ``Nucleon Form Factors and Hidden Symmetry in Holographic QCD,''
  Phys.\ Rev.\  D {\bf 77}, 014030 (2008)
  [arXiv:0710.4615 [hep-ph]].
}

\lref\GrigoryanMY{
  H.~R.~Grigoryan and A.~V.~Radyushkin,
  ``Structure of Vector Mesons in Holographic Model with Linear Confinement,''
  Phys.\ Rev.\  D {\bf 76}, 095007 (2007)
  [arXiv:0706.1543 [hep-ph]].
}

\lref\BayonaBG{
  C.~A.~B.~Bayona, H.~Boschi-Filho, M.~Ihl and M.~A.~C.~Torres,
  ``Pion and Vector Meson Form Factors in the Kuperstein-Sonnenschein
  holographic model,''
  JHEP {\bf 1008}, 122 (2010)
  [arXiv:1006.2363 [hep-th]].
}

\lref\IntriligatorAU{
  K.~A.~Intriligator and N.~Seiberg,
  Nucl.\ Phys.\ Proc.\ Suppl.\  {\bf 45BC}, 1 (1996)
  [arXiv:hep-th/9509066].
}

\lref\SeibergPQ{
  N.~Seiberg,
  ``Electric - magnetic duality in supersymmetric nonAbelian gauge theories,''
  Nucl.\ Phys.\  B {\bf 435}, 129 (1995)
  [arXiv:hep-th/9411149].
}

\lref\BandoEJ{
  M.~Bando, T.~Kugo, S.~Uehara, K.~Yamawaki and T.~Yanagida,
  ``Is Rho Meson A Dynamical Gauge Boson Of Hidden Local Symmetry?,''
  Phys.\ Rev.\ Lett.\  {\bf 54}, 1215 (1985).
}

\lref\BandoBR{
  M.~Bando, T.~Kugo and K.~Yamawaki,
  ``Nonlinear Realization and Hidden Local Symmetries,''
  Phys.\ Rept.\  {\bf 164}, 217 (1988).
}

\lref\AharonyQS{
  O.~Aharony,
  ``Remarks on nonAbelian duality in N=1 supersymmetric gauge theories,''
  Phys.\ Lett.\  B {\bf 351}, 220 (1995)
  [arXiv:hep-th/9502013].
}

\lref\PoppitzTX{
  E.~Poppitz and L.~Randall,
  ``Low-energy Kahler potentials in supersymmetric gauge theories with (almost)
  flat directions,''
  Phys.\ Lett.\  B {\bf 336}, 402 (1994)
  [arXiv:hep-th/9407185].
}

\lref\iss{
  K.~A.~Intriligator, N.~Seiberg and D.~Shih,
  ``Dynamical SUSY breaking in meta-stable vacua,''
  JHEP {\bf 0604}, 021 (2006)
  [arXiv:hep-th/0602239].
}

\lref\EckerTE{
  G.~Ecker, J.~Gasser, A.~Pich and E.~de Rafael,
  ``The Role Of Resonances In Chiral Perturbation Theory,''
  Nucl.\ Phys.\  B {\bf 321}, 311 (1989).
}

\lref\EckerYG{
  G.~Ecker, J.~Gasser, H.~Leutwyler, A.~Pich and E.~de Rafael,
  ``Chiral Lagrangians for Massive Spin 1 Fields,''
  Phys.\ Lett.\  B {\bf 223}, 425 (1989).
}

\lref\DavisMZ{
  A.~C.~Davis, M.~Dine and N.~Seiberg,
  ``The Massless Limit Of Supersymmetric QCD,''
  Phys.\ Lett.\  B {\bf 125}, 487 (1983).
}

\lref\AffleckMF{
  I.~Affleck, M.~Dine and N.~Seiberg,
 ``Exponential Hierarchy From Dynamical Supersymmetry Breaking,''
  Phys.\ Lett.\  B {\bf 140}, 59 (1984).
}

\lref\SeibergBZ{
  N.~Seiberg,
  ``Exact Results On The Space Of Vacua Of Four-Dimensional Susy Gauge
 Theories,''
  Phys.\ Rev.\  D {\bf 49}, 6857 (1994)
  [arXiv:hep-th/9402044].
}

\lref\KampfYF{
  K.~Kampf, J.~Novotny and J.~Trnka,
  ``On different lagrangian formalisms for vector resonances within chiral
  perturbation theory,''
  Eur.\ Phys.\ J.\  C {\bf 50}, 385 (2007)
  [arXiv:hep-ph/0608051].
}

\lref\McGarrieQR{
  M.~McGarrie,
  ``General Gauge Mediation and Deconstruction,''
  arXiv:1009.0012 [hep-ph].
}

\lref\AuzziMB{
  R.~Auzzi and A.~Giveon,
  ``The sparticle spectrum in Minimal gaugino-Gauge Mediation,''
  arXiv:1009.1714 [hep-ph].
}

\lref\SudanoVT{
  M.~Sudano,
  ``General Gaugino Mediation,''
  arXiv:1009.2086 [hep-ph].
}

\lref\GreenWW{
  D.~Green, A.~Katz and Z.~Komargodski,
  ``Direct Gaugino Mediation,''
  arXiv:1008.2215 [hep-th].
}

\lref\KutasovSS{
  D.~Kutasov, A.~Schwimmer and N.~Seiberg,
  ``Chiral Rings, Singularity Theory and Electric-Magnetic Duality,''
  Nucl.\ Phys.\  B {\bf 459}, 455 (1996)
  [arXiv:hep-th/9510222].
}

\lref\KutasovNP{
  D.~Kutasov and A.~Schwimmer,
  ``On duality in supersymmetric Yang-Mills theory,''
  Phys.\ Lett.\  B {\bf 354}, 315 (1995)
  [arXiv:hep-th/9505004].
}

\lref\KutasovVE{
  D.~Kutasov,
  ``A Comment on duality in N=1 supersymmetric nonAbelian gauge theories,''
  Phys.\ Lett.\  B {\bf 351}, 230 (1995)
  [arXiv:hep-th/9503086].
}

\lref\AmatiFT{
  D.~Amati, K.~Konishi, Y.~Meurice, G.~C.~Rossi and G.~Veneziano,
  ``Nonperturbative Aspects in Supersymmetric Gauge Theories,''
  Phys.\ Rept.\  {\bf 162}, 169 (1988) and references therein.
}

\lref\AffleckMK{
  I.~Affleck, M.~Dine and N.~Seiberg,
  ``Dynamical Supersymmetry Breaking In Supersymmetric QCD,''
  Nucl.\ Phys.\  B {\bf 241}, 493 (1984).
}

\lref\NovikovIC{
  V.~A.~Novikov, M.~A.~Shifman, A.~I.~Vainshtein and V.~I.~Zakharov,
  ``Supersymmetric instanton calculus: Gauge theories with matter,''
  Nucl.\ Phys.\  B {\bf 260}, 157 (1985)
  [Yad.\ Fiz.\  {\bf 42}, 1499 (1985)].
}

\lref\ShifmanMV{
  M.~A.~Shifman and A.~I.~Vainshtein,
  ``Instantons versus supersymmetry: Fifteen years later,''
  arXiv:hep-th/9902018.
}

\lref\CallanSN{
  C.~G.~.~Callan, S.~R.~Coleman, J.~Wess and B.~Zumino,
  ``Structure of phenomenological Lagrangians. 2,''
  Phys.\ Rev.\  {\bf 177}, 2247 (1969).
}

\lref\SakuraiJU{
  J.~J.~Sakurai,
  ``Theory of strong interactions,''
  Annals Phys.\  {\bf 11}, 1 (1960).
}

\lref\MasjuanAY{
  P.~Masjuan and S.~Peris,
  ``A Rational Approach to Resonance Saturation in large-Nc QCD,''
  JHEP {\bf 0705}, 040 (2007)
  [arXiv:0704.1247 [hep-ph]].
}

\lref\HaradaZJ{
  M.~Harada and K.~Yamawaki,
  ``Conformal phase transition and fate of the hidden local symmetry in  large
  N(f) QCD,''
  Phys.\ Rev.\ Lett.\  {\bf 83}, 3374 (1999)
  [arXiv:hep-ph/9906445].
}

\lref\HaradaJX{
  M.~Harada and K.~Yamawaki,
  ``Hidden local symmetry at loop: A new perspective of composite gauge boson
  and chiral phase transition,''
  Phys.\ Rept.\  {\bf 381}, 1 (2003)
  [arXiv:hep-ph/0302103].
}

\lref\ShifmanKD{
  M.~Shifman and A.~Yung,
  ``Confinement in N=1 SQCD: One Step Beyond Seiberg's Duality,''
  Phys.\ Rev.\  D {\bf 76}, 045005 (2007)
  [arXiv:0705.3811 [hep-th]].
}

\lref\ShifmanMB{
  M.~Shifman and A.~Yung,
  ``Non-Abelian Duality and Confinement in N=2 Supersymmetric QCD,''
  Phys.\ Rev.\  D {\bf 79}, 125012 (2009)
  [arXiv:0904.1035 [hep-th]].
}

\lref\ErlichQH{
  J.~Erlich, E.~Katz, D.~T.~Son and M.~A.~Stephanov,
  ``QCD and a Holographic Model of Hadrons,''
  Phys.\ Rev.\ Lett.\  {\bf 95}, 261602 (2005)
  [arXiv:hep-ph/0501128].
}

\lref\DaRoldZS{
  L.~Da Rold and A.~Pomarol,
  ``Chiral symmetry breaking from five dimensional spaces,''
  Nucl.\ Phys.\  B {\bf 721}, 79 (2005)
  [arXiv:hep-ph/0501218].
}

\lref\BeccioliniFU{
  D.~Becciolini, M.~Redi and A.~Wulzer,
  ``AdS/QCD: The Relevance of the Geometry,''
  JHEP {\bf 1001}, 074 (2010)
  [arXiv:0906.4562 [hep-ph]].
}

\lref\PomarolAA{
  A.~Pomarol and A.~Wulzer,
  ``Baryon Physics in Holographic QCD,''
  Nucl.\ Phys.\  B {\bf 809}, 347 (2009)
  [arXiv:0807.0316 [hep-ph]].
}


\Title{
} {\vbox{\centerline{Vector Mesons and an Interpretation of
Seiberg Duality} }}
\medskip

\centerline{\it Zohar Komargodski }
\bigskip
\centerline{School of Natural Sciences}
\centerline{Institute for Advanced Study}
\centerline{Einstein Drive, Princeton, NJ 08540}

\smallskip

\vglue .3cm

\bigskip
\noindent

We interpret the dynamics of Supersymmetric QCD (SQCD) in terms of
ideas familiar from the hadronic world. Some mysterious properties
of the supersymmetric theory, such as the emergent magnetic gauge
symmetry, are shown to have analogs in QCD. On the other hand,
several phenomenological concepts, such as ``hidden local
symmetry'' and ``vector meson dominance,''  are shown to be
rigorously realized in SQCD. These
considerations suggest a relation between the flavor symmetry group and the emergent gauge fields in theories with a weakly coupled dual description.
\Date{10/2010}

\newsec{Introduction and Summary}

The physics of hadrons has been a topic of intense study for
decades. Various theoretical insights have been instrumental in
explaining some of the conundrums of the hadronic world. Perhaps
the most prominent tool is the chiral limit of QCD. If the masses
of the up, down, and strange quarks are set to zero, the
underlying theory has an $SU(3)_L\times SU(3)_R$ global symmetry
which is spontaneously broken to $SU(3)_{diag}$ in the QCD vacuum.
Since in the real world the masses of these quarks are small
compared to the strong coupling scale,\foot{The approximation of
vanishing strange quark mass may seem dubious, but it works pretty
well in several circumstances.} the $SU(3)_L\times
SU(3)_R\rightarrow SU(3)_{diag}$ symmetry breaking pattern
dictates the existence of~$8$ light pseudo-scalars in the adjoint
of $SU(3)_{diag}$. These are identified with the familiar pions,
kaons, and eta.\foot{Several ideas which can be made precise at
large $N_c$ allow to include the eta' in this picture as well, as
the Goldstone boson of the axial symmetry~\WittenVV.} The
spontaneously broken symmetries are realized nonlinearly, fixing
the interactions of these pseudo-scalars uniquely at the two
derivative level. See~\WeinbergKR, along with references therein,
for a systematic exposition of these ideas.

The next hadrons one encounters are the vector mesons, consisting
of the rho mesons (with masses around $770$ MeV) and their
$SU(3)_{diag}$ partners. The analysis of the chiral limit does not
place stringent constraints on their dynamics. However, there are
strong phenomenological hints of an underlying structure. First of
all, to a good approximation, the rho mesons couple equally strongly to pions and nucleons.\foot{However, the coupling of the rho mesons to themselves is still unknown.} Secondly, many processes are saturated by
vector meson exchanges. This is usually referred to as ``vector
meson dominance''~\SakuraiJU. Finally, basic parameters associated to the
vector mesons (approximately) satisfy curious empirical relations.
Perhaps the most striking one~\refs{\KawarabayashiKD,\RiazuddinSW}
is $m_{\rho}^2=2g_{\rho\pi\pi}^2f_\pi^2$, where $g_{\rho\pi\pi}$
is the coupling of the rho meson to two pions.

One can attempt to account for these properties by imagining that
the rho mesons (and their $SU(3)_{diag}$ friends) are the gauge
fields of  a hidden local $[SU(3)]$ gauge
symmetry~\refs{\BandoEJ,\BandoBR}. (Of course, the hidden local
symmetry $[SU(3)]$ must be higgsed to reproduce the physical
nonzero masses of the rho mesons.) Coupling universality may be
readily explained by the universality of gauge interactions. The
relation $m_{\rho}^2=2g_{\rho\pi\pi}^2f_\pi^2$ can be interpreted
in terms of the usual formula $m_V^2\sim g^2v^2$, suggesting that
the hidden $[SU(3)]$ symmetry is higgsed at the scale $f_\pi$.
Lastly, with a little more work, vector dominance can be
reproduced too. This is without doubt a successful
phenomenological description of vector mesons. (Another interesting point of view on the subject is given in~\MasjuanAY.)

It is appropriate to question the uniqueness (and validity) of the
picture outlined above. We know that higgsed gauge symmetries are
not physical. Any other way to describe massive spin one particles
must yield the same results, for example, one can decide to
describe the spin one particles via antisymmetric
tensors~\refs{\EckerTE\EckerYG\BirseHD-\KampfYF}. To construct a map
between the various descriptions one must include high dimension
operators systematically. Indeed, in the presence of unsuppressed
high dimension operators there are absolutely no unique
predictions stemming from the existence of a hidden local
symmetry. The surprise that QCD offers is that the {\it minimal}
two derivative Lagrangian based on a hidden local symmetry is
capable of reproducing a host of phenomena with acceptable
precision. We find this rather astounding given that we are not
aware of any small parameter that might suppress the high
dimension operators.\foot{Another striking success of a phenomenological approach to QCD is Weinberg's original application of his sum rules~\WeinbergKJ\ (see also~\refs{\BernardCD,\DonoghueXB} and the sum rules developed in~\ShifmanBX.)}

Such arguments tempt one to conclude that the hidden local
symmetry is, in some sense, ``real.'' To prove this one would
first need to show that QCD is continuously connected to a theory
in which the rho mesons are massless (or just very light for some
reason) and that the parameter connecting these theories somehow
prevents large corrections from high dimension operators. In this
paper we do not attempt to shed any light on this possibility, but
we will address in detail a closely related, preliminary,
question: {\it Are there theories in which analogs of the rho
mesons are light?} We will argue that supersymmetric QCD in some
region of its parameter space is such a theory.

The salient features of the dynamics of SQCD have been understood
in a series of outstanding
insights~\refs{\DavisMZ\AffleckMK\NovikovIC\AmatiFT\ShifmanMV\SeibergBZ-\SeibergPQ},
reviewed in~\IntriligatorAU. A lightning review of these results
goes as follows. Consider an $[SU(N_c)]$ gauge theory (assuming
$N_c>2$) with $N_f$ flavors. For $N_f=0$ there are $N_c$ vacua,
each of which has a gap of order of the strong coupling scale. For
$0<N_f<N_c$ the theory has no supersymmetric vacua at finite
distance in field space and runs away to infinity. For $N_f=N_c$
and $N_f=N_c+1$ there are moduli spaces of vacua, and the weakly
coupled low energy excitations are identified with the gauge
invariant operators of the original theory (in other words, the
original baryons and mesons). The next phase one encounters is
$N_c+1<N_f<{3\over 2}N_c$. Again, there is a moduli space of
vacua. In particular, there is a supersymmetric vacuum at the
origin, where all the expectation values vanish. However, the
infra-red fluctuations cannot be just the baryons and mesons of
the original theory; anomaly matching forbids that.
Seiberg~\SeibergPQ\ has managed to identify the low energy
fluctuations. These are weakly coupled fields with canonical
kinetic terms, consisting of an $[SU(N_f-N_c)]$ IR-free gauge
theory with $N_f$ magnetic quarks, and, in addition, a
gauge-singlet matrix in the bi-fundamental representation of the
flavor group. Obviously, these degrees of freedom are very
different from the original variables. The origin of these IR
degrees of freedom is shrouded in mystery.

We will see that these magnetic gauge fields associated to
$[SU(N_f-N_c)]$ are identified naturally with the familiar rho
mesons. Furthermore, the magnetic quarks enter into the
story naturally as well. Amusingly, it turns out that SQCD also satisfies
vector meson dominance and several other benchmark properties of
vector mesons in QCD. Since the rho mesons of SQCD are light
(actually massless at the origin of the moduli space) the idea of
a hidden local symmetry is on theoretically firm footing, in contrast to
the case of QCD. We therefore illuminate some of the mysterious
features regarding the dynamics of SQCD in terms of ideas familiar
from nuclear physics. SQCD provides an
example in which these rough ideas are in fact precise.\foot{Other possible aspects of the similarity between the hidden local symmetry paradigm and Seiberg duality were suggested in~\refs{\HaradaZJ,\HaradaJX}.}

The outline of the paper is as follows. In section~2 we briefly
review the ordinary theory of pions and introduce vector mesons.
Our discussion of these topics is essentially a summary of known
results, with incidental original observations. In section~3 we
discuss supersymmetric QCD and provide evidence for our main
claim. An appendix contains some comments on vector mesons in the
large $N_c$ limit of QCD.

\newsec{Pions and Vector Mesons}
\subsec{Basics of the Theory of Pions} In this subsection we
recall how to write Lagrangians for theories with nonlinearly
realized symmetries, and we prepare the grounds for the inclusion
of vector mesons. For simplicity, we will only discuss the chiral
Lagrangian for the breaking of $SU(2)_L\times
SU(2)_R\hookrightarrow SU(2)_{diag}$.

The spectrum consists of three pions which are conveniently
assembled into a special unitary matrix\foot{Our conventions are
$$T^1=\left(\matrix{0 & 1 \cr 1 & 0}\right)~,\qquad
T^2=\left(\matrix{0 & -i \cr i & 0}\right)~,\qquad
T^3=\left(\matrix{1 & 0 \cr 0 & -1}\right)~.$$ The following two
identities are often useful (we define $\epsilon^{123}=1$)
$$T^aT^b=\delta^{ab}+i\epsilon^{abc}T^c~,$$
$$e^{i\pi^aT^a}=\cos\left(\sqrt{\vec\pi^2}\right)+i{\pi^aT^a\over\sqrt{\vec\pi^2}}\sin\left(\sqrt{\vec\pi^2}
\right)~.$$ } \eqn\stwomatrix{U=e^{i\pi^a T^a}~.} The
$SU(2)_L\times SU(2)_R$ symmetry is realized by acting on the
matrix $U$ simply as $U'=g_L U g_R^\dagger$. There is a unique
invariant Lagrangian at the two derivative level
\eqn\action{\CL={1\over 4} f_\pi^2Tr\left(\del_\mu U\del^\mu
U^\dagger\right)~.} Note that the diagonal symmetry with $g_R=g_L$
acts linearly on the pions while the axial transformations do not.
We can expand the Lagrangian in the number of pions. The first two
terms take the form \eqn\derivative{\CL=\half
f_\pi^2\left((\del\vec\pi)^2-\half\vec\pi^2(\del\vec\pi)^2+\cdots\right)~.}

There is another, equivalent, description of this system that will
be more useful for us. The idea is that we can factorize the
matrix $U(x)$ in terms of two special unitary matrices $\xi_L$ and
$\xi_R$ as follows \eqn\factor{U(x)=\xi_L(x)\xi_R^\dagger(x)~.}
This factorization is redundant. The theory has gauge invariance
which allows us to redefine $\xi_L\rightarrow \xi_L h(x)$,
$\xi_R\rightarrow \xi_R h(x)$ with any special unitary matrix
$h(x)$. The global $SU(2)_L\times SU(2)_R$ symmetry transformation
laws are $\xi_L\rightarrow g_L\xi_L$, $\xi_R\rightarrow g_R\xi_R$.
One can rewrite the theory~\action\ in terms of these redundant
degrees of freedom as follows \eqn\lagnewvar{\CL=-{f_\pi^2\over
4}Tr\left[\left(\xi_L^\dagger\del_\mu\xi_L-\xi_R^\dagger\del_\mu\xi_R\right)^2\right]~.}
It is easy to check that this Lagrangian is gauge invariant and it
is also invariant under global symmetry transformations~\CallanSN.

The physical properties of the pions can be calculated by fixing a
gauge. For example, we can choose to fix a gauge in which
$\xi_L=\xi_R^\dagger$. Global symmetry transformations take us out
of this gauge, but we can always reinstate our gauge choice by an
accompanying gauge transformation.

It is useful to think of this theory in the following language.
The $[SU(2)]$ gauge symmetry endows the model with a quiver-like
structure $SU(2)_L\times [SU(2)]\times SU(2)_R$, where $\xi_L$ is
in the bi-fundamental of $SU(2)_L\times [SU(2)]$ and $\xi_R$ is in
the bi-fundamental of $[SU(2)]\times SU(2)_R$. The vacua of this
theory are parametrized by constant matrices $\xi_L$, $\xi_R$,
modulo gauge transformations. So we can always choose
$\xi_L=1$ and $\xi_R$ is a general special unitary matrix. This
VEV for $\xi_L$ breaks the gauge symmetry but a diagonal flavor
symmetry coming from a mixture of the global transformations in
$SU(2)_L$ and (global) gauge transformations in $[SU(2)]$ remains.
(This pattern of flavor generators mixing with gauge generators
will be a recurring theme.) Then, the VEV for $\xi_R$ breaks the
flavor symmetry to $SU(2)_{diag}$. Note that so far in this model
there is a gauge symmetry but no gauge fields.

\subsec{Adding Gauge Fields}

The second version of the theory of pions~\lagnewvar\ has a
redundancy but no gauge fields associated to this redundancy.
Consider adding such a triplet of real vector fields $\rho_\mu^a$
transforming as usual \eqn\vectrans{\rho_\mu\equiv
\rho_\mu^aT^a\rightarrow h^\dagger \rho_\mu^aT^ah+ih^\dagger \del_\mu
h~.} We can construct two natural objects transforming
homogeneously under~\vectrans\
\eqn\twonatobj{\rho_\mu^L=\rho_\mu-i\xi_L^\dagger\del_\mu\xi_L~,\qquad
\rho^R_\mu=\rho_\mu-i\xi_R^\dagger\del_\mu\xi_R~.} At the two
derivative level the most general Lagrangian symmetric under
$L\leftrightarrow R$ can be written as
\eqn\leadingterms{\CL=-{1\over
g^2}(F_{\mu\nu}^{a})^2+{f_\pi^2\over
4}Tr\left[\left(\rho_\mu^L-\rho_\mu^R\right)^2\right]+a{f_\pi^2\over
4}Tr\left[\left(\rho_\mu^L+\rho_\mu^R\right)^2\right]~.} So far, $a,g$ are
undetermined real parameters. We will see that this theory
includes a massive spin one particle with mass of order $gf_\pi$,
so we should discuss its regime of validity.

First of all, an effective action for massive particles is subtle since, using the equations of motion, operators with different numbers of derivatives can mix. Secondly, for massive particles, one needs a small parameter that justifies  truncating the effective action to include finitely many terms. Therefore, for now, to make sense of the effective theory~\leadingterms\ we will assume that the gauge coupling $g$ is parametrically small. (Conversely, when the spin 1 fields are parametrically light, one must use gauge theories for consistency.) Eventually, we would like the massive spin one particles in the theory above to be identified with the rho mesons of QCD.\foot{We are ignoring the axial vector mesons for simplicity.} In nature, the gauge coupling of the rho mesons is by no means small. In spite of this, the theory~\leadingterms\ reproduces some of the properties of QCD remarkably well.

Note that $a=1$ is special in~\leadingterms. In this case $\xi_L$ interacts with $\xi_R$ only through gauge fields. As a consequence, when $g=0$, the global symmetry is enhanced due to the global gauge transformations to $SU(2)_L\times SU(2)^2\times SU(2)_R$. This symmetry argument has led Georgi~\refs{\GeorgiGP,\GeorgiXY} to propose the importance of $a=1$.
It is an inspiring idea, but unfortunately, the resulting theory does not seem to describe QCD. We will see that QCD is best described by a different, also special, value of $a$.

Denoting $\xi_L=e^{i\pi_L^aT^a}$, $\xi_R=e^{i\pi_R^aT^a}$ and expanding~\leadingterms\ to quadratic order we get
\eqn\lagexp{\CL=-{1\over g^2}(F_{\mu\nu}^a)^2+{f_\pi^2\over 2}\left(\del_\mu(\pi_L^a-\pi_R^a)\right)^2+{af_\pi^2\over 2}\left(\del_\mu(\pi_L^a+\pi_R^a)+2\rho_\mu^a\right)^2+\cdots~.}
In order to find the physical spectrum we pick unitary gauge $\pi_L=-\pi_R\equiv\pi$. We find a triplet of massless pions and a massive gauge field with mass
\eqn\mass{m_\rho^2=ag^2f_\pi^2~.}
More generally, the interesting terms in the interacting Lagrangian in this unitary gauge can be easily calculated (and simplified by using the free equations of motion and integration by parts)
\eqn\Lagii{\CL=-{1\over g^2}(F_{\mu\nu}^a)^2+2f_\pi^2\left((\del_\mu\pi)^2-2\pi^2(\del_\mu\pi)^2\right)+2af_\pi^2\left(\epsilon^{abc}\pi^a\del_\mu\pi^b+\rho_\mu^c\right)^2+\cdots~.}
The equation of motion of the massive gauge field sets it at low energies to $-\epsilon^{abc}\pi^a\del_\mu\pi^b$. Plugging this into~\Lagii\ and comparing with~\derivative\ we verify that $f_\pi$ is indeed correctly identified as the pion decay constant.

We would like to evaluate the global $SU(2)_{diag}$ currents $(\CJ_{diag}^a)_\mu$. This calculation is best done before fixing a gauge, in order to guarantee that our expression in unitary gauge descends from a gauge invariant operator. The complete gauge invariant expression for the conserved current is
\eqn\conserved{(\CJ_{diag}^a)_\mu={f_\pi^2\over 4}Tr\left[\left(\rho^L_\mu-\rho^R_\mu\right)\left(\xi_L^\dagger T^a\xi_L-\xi_R^\dagger T^a\xi_R\right)\right]+{af_\pi^2\over 4}Tr\left[\left(\rho^L_\mu+\rho^R_\mu\right)\left(\xi_L^\dagger T^a\xi_L+\xi_R^\dagger T^a\xi_R\right)\right]~.}
In unitary gauge the expression above becomes \eqn\current{(\CJ_{diag}^a)_\mu=2af_\pi^2\rho_\mu^a+2f_\pi^2(a-2)\epsilon^{abc}\pi^b\del_\mu\pi^c+{\rm three \ particles}+\cdots~.}

The immediate lessons from this formula are twofold. First, we see
that upon setting $a=2$, the coefficient of the second term
vanishes. The fact that $a=2$ is special is inconspicuous in the
original Lagrangian~\leadingterms. However, we will see that $a=2$
is actually the value which best describes the phenomenology of
QCD. A second corollary is that there is a general relation
between the physical coupling of the rho meson to pions
$g_{\rho\pi\pi}\equiv \half ga$, and the amplitude with which the
current creates a photon $g_{\rho\gamma}\equiv gaf_\pi^2$,
\eqn\relation{g_{\rho\gamma}=2g_{\rho\pi\pi}f_\pi^2~.} Note that
the unknown parameters $a$, $g$ cancel from this relation. Indeed,
this relation was recognized very early on via current algebra
techniques by KSFR~\refs{\KawarabayashiKD,\RiazuddinSW} (also
reviewed, for example, in~\BirseHD). To derive this relation we
assumed that $g$ is very small, but one can brazenly test this
relation in QCD. The agreement is about $10\%$, which is
remarkable.

The next two lessons to draw from~\Lagii,\current\ are a little less straightforward. One should study the electromagnetic form factor of the charged pion. We prepare a charged pion with four-momentum $p$ at early times which is then struck by an off-shell photon. The pion eventually leaves the interaction point with four-momentum $p'$. To calculate the result of this process we evaluate the electromagnetic current matrix element (denoting $q=p'-p$)
\eqn\pionformi{\langle\pi(p)\left|J^{QED}_\mu(0)\right|\pi(p')\rangle=(p+p')_\mu F(q^2)~.}

Equation~\pionformi\ follows from the masslessness of the pion and current conservation. From~\current\ we see that at tree-level the process has two components: a direct contact term between the current and the pions or an emission of a rho meson which propagates as a virtual particle and then, using the vertex $~\sim\epsilon^{abc}\rho^{\mu a}\pi^b\del_\mu\pi^c$, hits the target. These two contributions have a different analytic form. One more thing we know is that as a consequence of the fact that the pion has charge 1, in the deep infrared the form factor is independent of $a$ and satisfies $F(q^2=0)=1$.

We find \eqn\formfactor{F(q^2)= (1-\half a)+{\half am_\rho^2\over m_\rho^2-q^2}~.}
We can now start to appreciate why the special point $a=2$ is called the point of ``vector dominance.'' It is because the effect of scattering a photon on a pion target is fully accounted for by a $\rho$ exchange. Said in other words, the photon and rho gauge boson are maximally mixed. The fact that this process is saturated by  $\rho$ meson exchange also implies that a series of other processes is controlled by $\rho$ mesons,\foot{Another miracle that happens for $a=2$ is that the coupling of three rho mesons is of the same strength as the coupling to two pions. This is the famous coupling universality hypothesis in QCD.} but we will not discuss this here.
Note that choosing $a=2$, the general relation~\mass\ becomes in terms of $g_{\rho\pi\pi}$
\eqn\ksfrtwo{m_\rho^2=2g_{\rho\pi\pi}^2f_\pi^2~.}
This is another relation that is obeyed by the real world (at a level of around $5\%$).

One can provide a rationale for choosing $a=2$ by an argument akin to the Weinberg sum rules.
Asymptotic freedom tells us that in the deep Euclidean region the form factor drops like a power \eqn\asf{\lim_{q^2\rightarrow -\infty}F(q^2)\sim {1\over q^2}~.}
Therefore by extending the form factor to an analytic function in the complex $q^2$ plane and integrating $F(q^2)/q^2$ over a large contour we get zero. This means that the integral of the imaginary part over the time-like domain is zero.
Following Weinberg's original derivation of his sum rules~\WeinbergKJ, we ignore all the contributions besides those associated to the light resonances. Then, considering some contour $\gamma$ that encircles the origin and the rho meson pole we get
\eqn\cauchy{\int_\gamma d(q^2){F(q^2)\over q^2}=0~.}
Since $F(0)=1$ and the residue at $q^2=m_{\rho}^2$ follows from~\formfactor,
we arrive at
\eqn\vectordominance{a=2~.}
One has thus established vector meson dominance. (Vector dominance is often regarded as an input. Our point of view is that it is not a random fact about the hadronic world, rather, under some circumstances it could have been predicted by sum rules, in the same way that Weinberg predicted the axial vector mesons.)

We can even {\it estimate} the numerical value of the dimensionless parameter $g_{\rho\pi\pi}$. For this we have to consider the function $F(q^2)$ itself, as function of a complex variable. The coefficient of $1/q^2$ in~\asf\ can be calculated by explicit Feynman diagrams. In fact, the coefficient logarithmically runs to zero at infinitely large energies. Since this decay of the coefficient is only logarithmic, to correctly utilize the fact that \eqn\cauchyi{\int_\infty d(q^2)F(q^2)=0~,}
it is most convenient to read out the coefficient of $1/q^2$ in a region where asymptotic freedom already dominates but the logarithmic running has not been substantial. The experimental results~\refs{\HornTM,\HuberID} suggest that Bjorken scaling is approximately true already at a few GeV and the coefficient of $1/q^2$ is around $0.45$ GeV$^2$. This is again related to a sum over resonances by the Cauchy theorem. The low energy contribution to this integral comes from the residue of the pole at the $\rho$ meson mass. The contour argument relates these two residues given that we neglect the heavy mesons (and multi-particle states). This, together with the previous result $a=2$, allows us to conclude that $g_{\rho\pi\pi}\sim 5.1$. The correct value is around $6$. This estimate is as impressively successful as many of the results obtained via sum rules.

\subsec{Summary}

Let us summarize some of the important points we discussed in this section. It will be important to keep these in mind for our discussion of supersymmetric QCD.

\item{1.} Vector mesons are included in the chiral Lagrangian by splitting the pion field into two redundant pieces and adding gauge fields for this redundancy. Then a sequence of symmetry breaking phenomena takes place. First, the gauge symmetry is broken and the full $SU(2)_L\times SU(2)_R$ flavor symmetry survives as a linear combination of flavor generators and gauge generators. Subsequently, the flavor symmetry is further broken to $SU(2)_{diag}$.
\item{2.} This description makes physical sense only for small values of the gauge coupling (with fixed $f_\pi)$, but it seems to describe many important properties of nature in spite of the fact that the gauge coupling of the $\rho$ mesons in nature is pretty large.
\item{3.} The $\rho$ mesons can be created from the vacuum by the action of unbroken flavor symmetry generators.
\item{4.} In the framework of the two derivative effective field
theory, the relation~\relation\ follows. If one further applies a
sum rule in the usual way, one also finds vector meson dominance
(meaning that the $\rho$ mesons can fully account for various
physical effects such as the form factor we investigated) and the
second KSFR relation~\ksfrtwo. Note that if, in some sense,
effects from heavier states in QCD are small, then treating the
theory as if higher derivative terms are less important than the
leading ones, and applying the sum rule, would both be justified.
The experimental success of the results obtained suggests that the
effects of heavier states are indeed small.
\item{5.} There is no Higgs field that accompanies the massive $\rho$ mesons. However, in theories with light rho mesons, we surely expect to find Higgs fields. We will see in the next section that this is indeed what happens in supersymmetric QCD.

\newsec{Supersymmetric Quantum Chromodynamics}
We consider $SU(N_c)$ gauge theory with $N_f$ flavors $Q^i,\tilde
Q_i$, $i=1...N_f$. Our interest lies mostly in the IR-free non-abelian phase of SQCD, $N_c+1<N_f<{3\over
2}N_c$. The symmetry group is
$SU(N_f)_L\times SU(N_f)_R\times U(1)_B\times U(1)_R$, where the non-anomalous R-symmetry is $U(1)_R(Q)=U(1)_R(\tilde Q)=1-N_c/N_f$.

At energies much below the strong coupling scale, this theory flows to the Seiberg dual~\SeibergPQ\ $SU(N_f-N_c)$ IR-free gauge theory with $N_f$
magnetic quarks $q_i,\tilde q^i$ and a gauge-singlet
matrix $M^{i}_{j}$ in the bi-fundamental representation of the
flavor group. Obviously, these degrees of freedom are very
different from the original variables, but the vacua
agree upon introducing the superpotential \eqn\superdual{W=\tilde
q^jM^i_jq_i~.} In addition, the deformations of the two theories agree. (And the anomalies of course match.)

For small VEVs, these ``magnetic'' fields have a canonical K\"ahler
potential, albeit with an unknown normalization.
Neither the $SU(N_f-N_c)$ magnetic gauge fields nor the magnetic
quarks appear as well defined local operators in the UV. This must be so because they are charged under a hidden local symmetry group, so they are not gauge invariant.

Our goal here is to establish a dictionary (or an analogy) between
ideas familiar in ordinary QCD and the low energy description of supersymmetric QCD. We will provide evidence for the claim that the gauge fields should be thought of as rho mesons and the magnetic quarks are analogous to $\xi_L,\xi_R$.
As we have already mentioned, we will see that many ideas that in QCD work well
phenomenologically but are hard to justify theoretically can, in
fact, be justified in the supersymmetric version. Another benefit
of taking this analogy seriously is that many hitherto mysterious
features of SQCD can be understood as cousins of familiar ideas
from QCD.

\subsec{On the Moduli Space}

At the origin of the moduli space of SQCD the magnetic gauge fields are
massless, so to test our proposal we need to move away (slightly)
from the origin. Let us consider the following
direction in moduli space \eqn\baryon{q\equiv
\left(\matrix{\chi_{(N_f-N_c)\times (N_f-N_c)} \cr
\varphi_{N_c\times (N_f-N_c)}}\right)=v\left(\matrix{1& 0 & ...& 0
\cr 0& 1 & ...&0\cr . & . &. &.\cr . & . & . & .\cr 0&0&...&1\cr 0
& 0&...&0\cr. & . & . & .\cr. & . & . & . }\right)~.} To study the
physics along this flat direction it is also convenient to
decompose the other magnetic quark and meson \eqn\moredecom{\tilde
q\equiv \left(\matrix{\tilde \chi_{(N_f-N_c)\times (N_f-N_c)}, &
\tilde \varphi_{ (N_f-N_c)\times N_c}}\right)~,\qquad
M\equiv\left(\matrix{X_{(N_f-N_c)\times (N_f-N_c)} & Y \cr \tilde
Y & Z_{N_c\times N_c}}\right)~.}

Along this flat direction, the symmetry is broken from the original global symmetry
$SU(N_f)_L\times SU(N_f)_R\times U(1)_B\times U(1)_R$ to
$SU(N_f-N_c)_L\times SU(N_c)_L\times SU(N_f)_R\times U(1)'_B\times U(1)'_R
$.\foot{The primes on the various $U(1)$ groups mean that
they survive by mixing with non-Abelian flavor generators and
perhaps among themselves.} Therefore, the breaking along this
baryonic branch is essentially of the form \eqn\breaking{SU(N_f)_L\hookrightarrow
SU(N_f-N_c)_L\times SU(N_c)_L~.} The massless particles consist of
 $2N_fN_c-2N_c^2+1$ Goldstone bosons (one of
them is just the expectation value $v$ itself) and the massless mesons $Y,Z$. The IR theory along this moduli space does not include massless gauge fields and it is therefore a simple IR-free nonlinear sigma model for the coset~\breaking\ coupled to the massless mesons.\foot{This sigma model also matches the one obtained from the electric theory, as
was described in~\AharonyQS.}

For small $v$ the magnetic dual variables allow us to describe the correct
light (but not necessarily massless) excitations around this flat
direction. The dual magnetic gauge theory
$[SU(N_f-N_c)]$ is completely higgsed, hence, for small $v$, there are light
massive gauge fields along this flat direction with mass scaling like $\sim
gv$. In addition, the mesons $X,\tilde Y$ as well as the magnetic quarks $\tilde q$ are all massive (but light) with mass of order $v$.

Already at this stage we see some {\it superficial} hints for the
correspondence we are proposing. First of all, the flavor symmetry
$SU(N_f-N_c)_L$ survives at very low energies because in the
magnetic description it mixes with global gauge
transformations.\foot{This mixing, which occurs naturally in SQCD,
has been recently used for phenomenological purposes~\GreenWW.
Some related discussions can be found
in~\refs{\McGarrieQR\AuzziMB-\SudanoVT}. It would be nice to see
if there are connections between our approach and the analysis
in~\refs{\ShifmanKD,\ShifmanMB}, where this color-flavor locking
phenomenon played a major role.} In other words, the Goldstone
bosons for the breaking~\breaking\ are the magnetic quarks
$\varphi_i^c$. The index $c$ transforms in the fundamental
representation of the unbroken $SU(N_f-N_c)_L$ flavor symmetry,
but it really descends from a gauge index at higher energies. Yet
another way to say the same thing is that given the nonlinear
sigma model for $SU(N_f)_L/ \left(SU(N_f-N_c)_L\times
SU(N_c)_L\right)$ the way we could reintroduce the magnetic gauge
fields into this theory is by promoting the redundant
$SU(N_f-N_c)_L$ transformations into a local symmetry and then
adding spin one particles with kinetic terms. This is precisely
the way we introduced the rho mesons into the pion Lagrangian in
section~2. We see that the magnetic quarks are also natural
players in the story, they are the redundant variables we add to
allow for a local symmetry, hence, they are analogous to the
$\xi_L,\xi_R$ degrees of freedom in the theory of the previous
section. (One difference is that SQCD also contains the Higgs
fields associated to global symmetry breaking. As mentioned in
subsection~2.3, this must have been the case since in SQCD a limit
in which the global symmetry is restored exists.)

So far these are merely intuitive similarities, but they can be
made precise by studying the global symmetry currents of the
theory~\superdual. Consider the $SU(N_f-N_c)_L$ global symmetry
current superfields. We can attempt to write them in terms of the
magnetic dual variables. However, since we do not know the
normalizations of the (canonical) kinetic terms, there are some
order one numbers that we cannot control (we do know their signs
though). This will not affect our discussion, and we will
henceforth simply suppress these incalculable (positive) numbers. The
expression for the currents then takes the form ($V\equiv V^aT^a$
is the magnetic vector superfield)
\eqn\currentsup{\CJ_{SU(N_f-N_c)_L}^a=\chi^c_{j}
(e^V)_c^d(\chi^\dagger)_d^{i}(T^a)^j_i-X^i_l(X^\dagger)^l_j(T^a)^{j}_i-(\tilde
Y)^i_l(\tilde Y^\dagger)^l_j(T^a)^j_i~.} The  meaning of this
expression is most transparent once we fix a unitary gauge for the
magnetic group. Unitary gauge means that we set all the quadratic
terms mixing the gauge field and the matter fluctuations to zero.
This is achieved by \eqn\gaugefix{\forall a. \ \
\langle\chi^c_{i}\rangle (T^a)_c^d(\delta\chi^\dagger)_d^{i}=0~.}
(The general theory of such unitary gauges has been developed
in~\PoppitzTX.) In our case,~\baryon,
$\langle\chi^c_{i}\rangle=v\delta^c_{i}$ and unitary
gauge~\gaugefix\ therefore just means that $\delta\chi^c_i\sim
\delta^c_i$. In other words, the only physical degree of freedom
in the $\chi$ magnetic quarks is the overall scale $v$, which is
one of the complex Goldstone bosons. We denote this special mode
by $\pi$, i.e.~$\chi^c_{i}=v\delta^c_{i}+\pi \delta^c_{i}$.

Evaluating the current~\currentsup\ in this gauge and dropping all
terms with more than two particles we find
\eqn\currentsupi{\CJ_{SU(N_f-N_c)_L}^a=v^2(e^V)_c^d(T^a)^c_d+v(\pi+\pi^\dagger)(e^V)_c^d(T^a)^c_d-{\rm
mesons}+\cdots~.} The meson terms that we suppressed are
identical to those in~\currentsup. The physical quantity we would
like to compute is the form factor of the Goldstone bosons
$\varphi$. We will be interested only in the tree-level
contributions. The terms quadratic in mesons surely cannot
contribute, since there is no way to draw a diagram at tree level.
Similarly, the second term in~\currentsupi\ plays no role. We
remain with the first term, in which only the piece linear in $V$
contributes at tree level. This gives
\eqn\formfacvar{F_\varphi(q^2)= {m_V^2\over m_V^2-q^2}~.} Since a term quadratic in $\varphi$ is absent from~\currentsupi, there is no constant piece in the form factor.

We see that not only is the identification between the Seiberg
dual gauge fields and the rho mesons manifest, SQCD also satisfies
vector dominance. So SQCD seems to sit at a point analogous to
$a=2$ in QCD (which, as we explained, is closest to describing
nature). That $a$ is equal $2$ in QCD has been motivated by
sum rules. However, in SQCD, the analogous result follows
rigorously from Seiberg duality.

Note that at energies below the massive vector boson the form
factor is $1$, as it should be. At such low energies we should
integrate out the massive vector field, and the massive fields
$\tilde q$, $X$, $\tilde Y$. Note that as far as $X$, $\tilde Y$
are concerned, they do not have quadratic mixing terms with light
fields and so the last two terms in~\currentsup\ cannot produce
terms quadratic in the light fields. We can thus ignore these
terms. What remains is to solve the equations of motion of the
massive vector fields. In our unitary gauge we get an expansion in
the number of light fields, starting with the following quadratic
term in the Goldstone bosons \eqn\solution{V^a\sim \varphi^c_i
(T^a)_c^d(\varphi^\dagger)_d^i~.} Plugging this back into the
expression for the current~\currentsup, we find that at low
energies the $SU(N_f-N_c)_L$ current is
\eqn\currentsupii{\CJ_{SU(N_f-N_c)_L}^a\sim \varphi^c_i
(T^a)_c^d(\varphi^\dagger)_d^i+{\rm three \ particles}+\cdots~.}
This is, of course, the expected result, since $\varphi$ transforms
linearly under $SU(N_f-N_c)_L$ transformations at very low
energies. The physical interpretation of what is happening here is
as follows. At energies above $v$ the fields $\varphi$ are not
charged under the global  $SU(N_f-N_c)_L$  symmetry
transformations and this is why they are absent from~\currentsup.
However, due to higgsing of the magnetic gauge symmetry and the
fact that $SU(N_f-N_c)_L$ is realized as a mixture of the original
flavor transformations and some global gauge transformations,
$\varphi$ is indeed charged under $SU(N_f-N_c)_L$ at low energies.
This is why there is a quadratic term in $\varphi$ in the
expression for the $SU(N_f-N_c)_L$ current at low
energies~\currentsupii. Since $\varphi$ has such an ``energy
dependent charge,'' vector meson dominance is realized in
earnest: at high energies the form factor is saturated by a gauge
field propagator, but at very low energies it is accounted for by
the structure-less charged particle $\varphi$.

We thus see that SQCD slightly deformed from the origin has a rich
structure that in some respects resembles QCD, especially in the
way vector mesons appear and the way they dominate physical
processes. The conclusion that the Seiberg dual gauge fields are
analogous to the $\rho$ mesons, and the magnetic quarks' role is
similar to those of $\xi_L$, $\xi_R$, is unavoidable. In addition
to the structural similarities, phenomenological properties such
as vector dominance, and relations analogous
to~\relation,\ksfrtwo, are satisfied.

One general comment is in order. In (the chiral limit of) QCD the full unbroken global symmetry creates vector mesons from the vacuum and they are pretty heavy. In SQCD, on the branch of moduli space we have studied here, the unbroken symmetry contains $SU(N_f-N_c)_L\times SU(N_c)_L$. We have shown that the currents of $SU(N_f-N_c)_L$ create light vector mesons from the vacuum, but $SU(N_c)_L$ has not played a major role. Indeed, we expect it to create heavy one particle states, therefore not visible in the Seiberg dual description; in the IR the $SU(N_c)_L$ currents only create states with more than one particle. Interestingly, the roles of $SU(N_f-N_c)_L$ and $SU(N_c)_L$ get reversed in a different region of parameter space.

Consider taking $N_f>3N_c$. In this case the electric theory is
not UV free but the magnetic theory~\superdual\ is. The flat
direction~\baryon\ still exists but now for small $v$ we cannot
analyze it in terms of the magnetic variables since they are
strongly coupled. The electric variables provide the weakly
coupled description, and the $[SU(N_c)]$ gauge theory is now the
hidden local symmetry which one encounters in the IR. It is not
hard to see that now the $SU(N_c)_L$ flavor symmetry survives in
the IR because of mixing with global gauge transformations and
that the currents of $SU(N_c)_L$ excite the light gauge bosons of
the hidden local symmetry.

We therefore see that the theory we are discussing has both light
and heavy rho mesons. The light rho mesons are created by some
global symmetry currents and the heavy ones are created by a
different set of global symmetry currents. We have no theoretical
control over the heavy rho mesons, but we can say a lot about the
parameterically light ones. A similar picture will emerge when we
study SUSY-breaking field configurations in the next subsection.

\subsec{Off the Moduli Space}

We would like to subject the identification we are proposing to
further tests. The idea is that the role of the dual gauge fields
as rho mesons {\it must} manifest even in small deformations of
the theory that may have a (small) nonzero vacuum energy density.

We consider again the free magnetic phase $N_c+1<N_f<{3\over
2}N_c$ and add a mass term to all the electric quarks
\eqn\masqcd{W_{electric}=mQ^i\tilde Q_i~.} The dynamics of this
theory for small field VEVs has unfolded only in recent years,
starting with~\refs{\iss}. The symmetry group of~\masqcd\ is
$SU(N_f)\times U(1)_B$. The description of the theory near the
origin is in terms of the Seiberg dual variables
\eqn\dualmsqcd{W_{magnetic}=\tilde q^jM^i_jq_i-\mu^2M^i_i~,} where
we have denoted $\mu^2=-m\Lambda$. The symmetry group of the theory contains $SU(N_f)$, but we definitely do not see $N_f^2-1$ gauge bosons in the IR, just $(N_f-N_c)^2-1$ of them. This apparent contradiction is resolved by carefully studying the dynamics of~\dualmsqcd.

The $F$-term equations for the meson field $q^jq_i-\mu^2 \delta^j_i=0$ cannot all be satisfied
because the rank of the first term is necessarily smaller than
$N_f$. This means that for small field VEVs there are no SUSY
vacua and the vacuum energy density is (at least) of order $\mu^4$
(we choose $m,\Lambda$ to be real without loss of generality). We
can trust these non-supersymmetric configurations as long as the typical VEVs and energy densities are much smaller than the cutoff. For this reason we focus on the regime $m\ll\Lambda$ (which also implies $\mu\ll\Lambda$).\foot{At field VEVs much
larger than $\sim\mu$ there are in fact supersymmetric vacua, but
they are of no interest to us here.}

Using the same notation as in~\baryon\ and~\moredecom, one finds
that the classical energy density is minimized by setting
$\chi_i^c=\mu\delta^c_i$, $\tilde\chi^i_c=\mu\delta_c^i$, while
all the other fields are set to zero. The symmetry is broken as
follows \eqn\symbre{SU(N_f)\times U(1)_B\hookrightarrow
SU(N_f-N_c)\times SU(N_c)\times U(1)'_B~.} We now see how the
apparent contradiction mentioned above is going to be resolved.
The crux of the matter is that, in the vacuum close to the origin,
the massive theory~\masqcd\ spontaneously breaks the $SU(N_f)$
global symmetry. Unlike supersymmetric vacua, in which the pattern
of symmetry breaking can be inferred at weak coupling, in this
case one must genuinely use the duality. There is no known way to
anticipate~\symbre\ from the electric description. On the other
hand, that the symmetry breaks in this way is absolutely necessary
for the consistency of the picture we are proposing.\foot{We thank
M.~Strassler for a helpful conversation on the subject.}

From here the story proceeds in parallel to the story in QCD. The
main difference being that some unbroken generators (those of
$SU(N_f-N_c)$) create light rho mesons from the vacuum, namely,
the Seiberg dual gauge fields. The other symmetry generators only
create massive particles, not visible at low energies. This is not
a contradiction, it only means that SQCD in this SUSY-breaking
vacuum has two distinct sets of rho mesons. Those associated to
$SU(N_f-N_c)$ are very light and can in fact be continuously taken
to be massless, while those of $SU(N_c)\times U(1)'_B$ are heavy
and cannot be analyzed analytically. Our main point is that the
existence of this set of arbitrarily light rho mesons allows,
among other things, to exhibit some phenomenological ideas from
QCD in a rigorous setup.

The Goldstone bosons bassociated with~\symbre\ are given by
$\varphi+\tilde \varphi^*$ and an additional real singlet mode
$Tr(Im(\chi-\tilde\chi))$. One also finds the so called
``pseudo-moduli,'' which parametrize classical non-compact flat
directions. In our case~\dualmsqcd, one-loop effects have been
shown~\iss\ to set these fields to zero.

As we have already implied, the currents of the unbroken $SU(N_f-N_c)$ symmetry are
candidates for creating the magnetic gauge fields from the vacuum. We
can write an explicit expression (up to an overall coefficient)
for the $SU(N_f-N_c)$ currents
\eqn\currentsup{\CJ_{SU(N_f-N_c)}^a=\chi^c_{j}
(e^V)_c^d(\chi^\dagger)_d^{i}(T^a)^j_i-
(\tilde \chi^\dagger)^c_{j}(e^{-V})_c^d\tilde\chi_d^{i}(T^a)^j_i+mesons~,}
where we have suppressed all the terms bilinear in mesons. The natural
choice of unitary gauge in this case is \eqn\gaugefixii{\forall a.
\ \ \langle\chi^c_{i}\rangle
(T^a)_c^d(\delta\chi^\dagger)_d^{i}-\langle\tilde\chi_d^{i}\rangle
(T^a)_c^d(\delta\tilde\chi^\dagger)^c_{i}=0~.} Plugging the VEVs
of $\chi$, $\tilde \chi$, we see that unitary gauge amounts to
setting \eqn\gaugefixiii{\delta\chi- \delta\tilde\chi\sim \unit~.}
Similarly to the discussion after~\gaugefix, the unit matrix
corresponds to some complexified transformations of the vacuum
expectation values. Since in this case the vacuum is not
supersymmetric, only the imaginary part of~\gaugefixiii\ is in fact
massless while the real part is one of the pseudo-moduli that
obtains a mass from radiative corrections.

In this gauge the current multiplet becomes
\eqn\currentsup{\CJ_{SU(N_f-N_c)}^a=\mu^2V^a +{\rm two \
particles}+\cdots~.} Hence, we see again that the unbroken
$SU(N_f-N_c)$ currents create the magnetic gauge
fields from the vacuum. Since at energy scales of order $\mu$ and above a quadratic
term $\sim \varphi T^a \varphi^\dagger$ is absent
from~\currentsup, the form factor for the Goldstone bosons
$\varphi+\tilde \varphi^*$ will be of the form~\formfacvar,
satisfying vector dominance. At low energies we can integrate out
the massive vector field and the current becomes quadratic in the
Goldstone bosons. The physics of this is identical to what we have found in the previous subsection.
This completes our analysis of the SUSY-breaking case, corroborating our proposal.

\subsec{Open Questions}

It would be nice to check whether the phenomena we found here are
general or not. To address this question, one would need to study
other examples, such as the orthogonal and symplectic cases. More
exotic examples, such as adjoint
SQCD~\refs{\KutasovVE\KutasovNP-\KutasovSS}, are also interesting
to study.

In general, it may be possible to identify the emergent gauge
bosons with rho vector mesons only if the number of emergent gauge
bosons is not larger than the number of flavor symmetry
generators. Thus, we obtain an inequality between two objects
which seem unrelated at first sight. Clearly, if vector-like
theories (in the IR-free phase) that violate this inequality
exist, they must be truly exotic.\foot{It is worth mentioning that
there may be supersymmetric theories in which the analogs of the
axial vector mesons and perhaps also the $\rho'$ mesons are
massless. In the case of $\rho'$, by definition, one global
symmetry current can create both the $\rho$ and $\rho'$ from the
vacuum. Naively, it seems that in this case the magnetic gauge
group would have to be a product group and so a more refined
version of our inequality would still hold. Needless to say, it
would be interesting to search for such theories and to
investigate their properties.} It would be interesting to conduct
a systematic survey of the known models, but here we will merely
check this inequality in adjoint SQCD. From our point of view, the
latter is a curious example because the rank of the emergent group
can be arbitrarily larger than the rank of the global symmetry
group, so if the inequality is to hold water, there must be a
nontrivial interplay between the location of the IR-free window,
the rank of the emergent gauge group, and the rank of the flavor
group.

Consider SQCD with gauge group $SU(N_c)$, $N_f$ electric quarks,
and an adjoint field $X$. We include the superpotential
\eqn\supadqcd{W=Tr(X^{k+1})~.} The matter fields in the dual
theory~\KutasovNP\ consist of a magnetic gauge group
$SU(kN_f-N_c)$, $N_f$ magnetic quarks, an adjoint field, and a
family of gauge singlets. Therefore, the dual theory is IR-free as
long as the beta function
\eqn\betadual{-\beta_{dual}=(2k-1)N_f-2N_c~,} is positive. In
addition, we must take $N_f\geq N_c/k$ (for $N_f<N_c/k$ the theory
has no vacuum). We conclude that the IR-free phase is realized for
\eqn\irfree{{N_c\over k}\leq N_f<{2N_c\over 2k-1}~.} To test our
inequality, it is sufficient to show that throughout this window
$kN_f-N_c<N_f$. Indeed, this is equivalent to satisfying
$N_f<N_c/(k-1)$, which is consistent with~\irfree.

This is just a preliminary necessary condition for our proposal to
be realized in adjoint SQCD. It would be nice to work out the
details in these and other theories.

\bigskip
\bigskip
 \noindent {\bf Acknowledgments:}
\nobreak We would like to thank N.~Arkani-Hamed, T.~Dumitrescu,
A.~Giveon, J.~Maldacena, S.~Peris, and N.~Seiberg for helpful
conversations. We would especially like to express our gratitude
to T.~Banks for sharing with us many of his insights. After this
work was completed, M.~Strassler kindly shared with us his
unpublished thoughts on the subject, and we would like to thank
him for several stimulating conversations. We gratefully
acknowledge support by DOE grant DE-FG02-90ER40542. Any opinions,
findings, and conclusions or recommendations expressed in this
material are those of the authors and do not necessarily reflect
the views of the National Science Foundation.

\appendix{A}{The Large $N_c$ Limit}

We would like to offer some comments on vector mesons in  QCD with many colors $N_c\rightarrow\infty$~\tHooftJZ. We start by recalling some well known facts about this limit (see~\WittenKH\ for more details).
At large $N_c$, the two point function of, say, flavor currents can be decomposed into a sum over all the meson states, which are exactly stable (and free) at infinite $N_c$. Since this is of order $N_c$, the amplitude to create a meson from the vacuum by the action of a current is $\sqrt{N_c}$. From this it follows that the pion decay constant scales like $f_\pi\sim \sqrt{N_c}$. The masses of these states do not scale with $N_c$. Similarly, by considering properties of three point functions, we learn that cubic interaction vertices are suppressed by $1/\sqrt{N_c}$. Importantly for us, this also shows that currents can create two particle states from the vacuum with amplitude of order 1.

Let us test this picture against the effective theory of $\rho$ mesons~\Lagii,\current. From the interaction of three $\rho$ mesons we immediately conclude that $g\sim {1\over\sqrt{N_c}}$. We then observe from the interaction of two pions and $\rho$ that the parameter $a$ must stay finite in the large $N_c$ limit. Note that this implies through~\mass\ that the mass of the $\rho$ meson stays finite for large $N_c$. From~\current\ we see that the amplitude to create a single rho meson from the vacuum scales like $gf_\pi^2\sim\sqrt{N_c}$, which is consistent with the general large $N_c$ rule. From~\current\ we also see that a pair of pions is created with amplitude of order~1, in accordance with large $N_c$ expectations. We therefore conclude that the theory of rho mesons is consistent with our expectations from large $N_c$, but large $N_c$ arguments do not fix $a$.

Let us now study the pion form factor, including the whole tower of vector mesons indexed by $n$. One can then imagine that the unbroken flavor current contains all these mesons and possibly a bilinear term $\pi\del\pi$ like in~\current. The form factor of the pion then takes the form
\eqn\formfactorlargN{F(q^2)=c+\sum_n{\kappa_n\over q^2-m_n^2}~.}
The constant $c$ comes from the possibility of the current to create a two pion state from the vacuum directly. The $\kappa_n$ are some dimension two coefficients which depend on the amplitude of the current to create the $n$th vector meson from the vacuum and also on the vertex which connects the $n$th vector meson with two pions. Thus, both $c,\kappa_n$ are finite for $N_c\rightarrow\infty$. We need to make sure that $F(0)=1$, which gives one relation among the infinitely many coefficients in~\formfactor. Assuming asymptotic freedom, one can write a sum rule $\int_\infty F(q^2)/q^2=0$ which then implies (to arrive at the formula below we use the fact that the residue at the origin is 1)
\eqn\sumrule{1+\sum_n{\kappa_n\over m_n^2}=0~.}
Together with the constraint $F(0)=1$ we find that $c=0$. Hence, under our assumptions, the form factor at large $N_c$ in theories which are asymptotically free must be of the resonance-saturated  form
\eqn\formasf{F(q^2)=\sum_n{\kappa_n\over q^2-m_n^2}~.}

This form is manifest in many examples,
e.g.~\refs{\SonET\HongSA\HongNP\GrigoryanMY\KweeDD\HongDQ\KweeNQ\RodriguezGomezZP-\BayonaBG}.
However, for such a theory to look close to nature one would like
to satisfy the relations~\relation,\ksfrtwo\ or at least satisfy
them approximately. This is not automatic and, for example, in the
AdS/QCD setup these relations are often not satisfied. (If, in
some sense, the contribution of the higher resonances is small,
the form factor will be dominated by the rho meson, as is the case
in QCD. This often occurs in AdS/QCD due to the oscillatory
behavior of the heavier KK wave functions.) See, for example,
\refs{\ErlichQH,\DaRoldZS} in which concrete cases were analyzed.
Even though the specific relations~\relation,\ksfrtwo\ do not
follow automatically from the AdS/QCD approach, when one fits many
quantities the agreement with the data is pretty good.
See~\BeccioliniFU\ for a recent detailed analysis and
also~\PomarolAA\ for another interesting aspect of vector meson
dominance analyzed in the context of AdS/QCD.

The result~\formasf\ (which we argued follows from consistency) is satisfied in models of AdS/QCD in an interesting way. It is translated to a certain completeness relation among the wave functions of KK states. Our point of view is that this is really just a consequence of the high energy behavior of the form factor, which is fixed by certain dimensions of operators in the UV CFT.\foot{We thank J.~Maldacena for a discussion that led to this claim.} (We have only discussed asymptotically free theories, but one could render the discussion general.)

\listrefs
\end